\documentclass[prl,twocolumn,showpacs,preprintnumbers]{revtex4}
\usepackage{graphicx}
\usepackage{dcolumn}
\usepackage{bm}
\usepackage{amsmath}

\begin{document}

\title{Wave Function and Pair Distribution Function of a Dilute Bose Gas}

\author{Bo-Bo Wei$^1$}
\author{Chen-Ning Yang$^{1,2}$}

\affiliation{$^1${Chinese University of Hong Kong, Hong Kong}
$^2${Tsinghua University, Beijing, China}}

\begin{abstract}

The wave function of a dilute hard sphere Bose gas at low
temperatures is discussed, emphasizing the formation of pairs. The
pair distribution function is calculated for two values of
$\sqrt{\rho a^3}$.

\end{abstract}

\date{\today}
\maketitle




The simplest nontrivial quantum mechanical many body problem that
is precisely definable and at the same time subject to
mathematical treatment is the dilute hard sphere Bose gas.  In
1947 Bogoliubov studied this problem \cite{NNBogo:47} and obtained
its excitation spectrum.  His method is now called
\cite{Khuang:87} the Bogoliubov transformation.  In 1957 Lee,
Huang and Yang made \cite{LHY}  a detailed study of the problem
and obtained its ground state energy in an asymptotic expansion:
\begin{equation}
\frac{E_0}{N}=(\frac{\hbar^2}{2m}) 4\pi a\rho
[1+\frac{128}{15\sqrt{\pi}}\sqrt{\rho a^3}+\ldots]
\end{equation}
where $N$ is the total number of particles, $\rho=N/\Omega$ is the
density of particles, and $a$ is the diameter of the sphere =
scattering length.

With incredibly advanced modern technologies this system is now
subject to experimental study.  Many beautiful results
\cite{LPit:98}  have been obtained, about equation (1) and about
the excitation spectrum. We want to point out in this brief note
that perhaps further interesting experimental exploration could be
made to further clarify properties of the ground state. Throughout
this note we follow the notation of reference [3].

\section{Wave Function for Ground State}

LHY had given a \emph{physical picture} of the ground state
$\Psi_0$.  It turns out to be a collection of $n$ pairs of
particles floating in a sea of $\textbf{k}=0$ particles.  Each
pair consists of two particles with momenta $\textbf{k}$ and
$-\textbf{k}$.  The pair is not a bound pair, but a
\emph{correlated pair} with a correlation length equals to
\begin{equation}
r_0=(8 \pi a\rho)^{-1/2}.
\end{equation}
$n=0,1,2...$, with $n = 0$ being dominant, $n=1$ the next
dominant, etc.  The average value of $n$ is
\begin{equation}
\langle n\rangle=\frac{4N}{3\sqrt{\pi}}\sqrt{\rho a^3}.
\end{equation}
Thus the probability of a particle being in $\textbf{k}\neq0$
state is $\frac{8}{3\sqrt{\pi}}\sqrt{\rho a^3}$. Within one
correlation length the number of particle is, on the average,
\begin{equation}
\sim \rho r_0^3=\frac{1}{\sqrt{8\pi}^3}\frac{1}{\sqrt{\rho
a^3}}\gg 1.
\end{equation}
But most of these would be in the $\textbf{k}=0$ state,  with only
a few particle having $\textbf{k}\neq0$:
\begin{equation}
\sim \frac{1}{\sqrt{8\pi}^3}\frac{1}{\sqrt{\rho
a^3}}\left[\frac{8}{3\sqrt{\pi}}\sqrt{\rho
a^3}\right]=\frac{1}{3\sqrt{8}\pi^2}.\nonumber
\end{equation}
In other words, within one correlation length, there are many
particles, but only a few with $\textbf{k}\neq 0$.  The
correlation therefore acts over a sea of particles almost all of
which have $\textbf{k}=0$. That characteristic is, we believe,
what was vaguely called momentum space ordering by London
\cite{FLondon} in 1954 for a superfluid system.

It is interesting to compare the present theory for Bosons with
the BCS theory for Fermions.  In both cases there is
\emph{formation of correlated pairs}.  But the mechanisms for the
formation are quite different.  The important common feature for
the two cases is the presence of ODLRO \cite{CNYang:62}.

\section{Pair Distribution Function}
\label{sec:sum}

The pair distribution function
\begin{eqnarray}
D(r_{12})&=&\rho^{-2}\langle\psi^{\dag}(\textbf{r}_1)\psi^{\dag}(\textbf{r}_2)\psi(\textbf{r}_2)\psi(\textbf{r}_1)\rangle
\end{eqnarray}
is an important physical quantity \emph{measurable} for many
liquid systems.  In this formula $\psi(\textbf{r})$ is the
annihilation operator in $\textbf{r}$ space.

The pair distribution function $D(r_{12})$ is also related to the
\emph{diagonal elements} of the reduced density matrix
\cite{CNYang:62}:
\begin{eqnarray}
D(r_{12})=\rho^{-2}\text{Trace}[\psi(\textbf{r}_2)\psi(\textbf{r}_1)\rho_N\psi^{\dag}(\textbf{r}_1)\psi^{\dag}(\textbf{r}_2)],\nonumber
\end{eqnarray}
where $\rho_N=\Psi_0 \Psi_0^{\dag}$.

The meaning of $D(r)$ is: given a particle at a point, the average
number of particles in $d^3\textbf{r}$ at a distance $r$ is $\rho
D(r)d^3\textbf{r}$. Thus $D(r)\rightarrow 1$ as $r\rightarrow
\infty$.

For the dilute Bose system under consideration, $D(r)$ has been
\cite{ass} computed in LHY:
\begin{eqnarray}
D(r)=[1+G(r)]^2+[1+F(r)]^2-1-2f[G(r)+F(r)]\nonumber \\
\hspace{2.8cm} [43'] \nonumber
\end{eqnarray}
where
\begin{eqnarray}
F(r)&=&\frac{1}{8\pi^3\rho}\int \frac{\alpha^2}{1-\alpha^2}e^{i\textbf{k}\cdot \textbf{r}}d^3\textbf{k},\nonumber\\
G(r)&=&-\frac{1}{8\pi^3\rho}\int
\frac{\alpha}{1-\alpha^2}e^{i\textbf{k}\cdot
\textbf{r}}d^3\textbf{k}. \hspace{2.4cm} [44]\nonumber
\end{eqnarray}

\begin{figure}
\begin{center}
\includegraphics[scale=0.8]{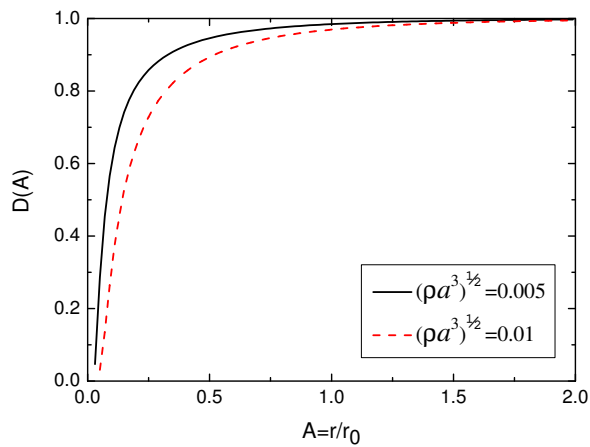}
\caption{\label{fig:epsart1}(color online)  Plot of $D(r/r_0)$.}
\end{center}
\end{figure}

The quantity $\alpha$ was defined in LHY by [38].  All equations
with [ ] refer to equations in LHY.  Converting the integrals
$\int d^3\textbf{k}$ into radial and angular parts yield
\begin{eqnarray}
F(r)&=&\frac{8}{A}\sqrt{\frac{2\rho a^3}{\pi}}I_1(A), \nonumber\\
G(r)&=&-\frac{8}{A}\sqrt{\frac{2\rho a^3}{\pi}}I_2(A),
\end{eqnarray}
\begin{equation}
\hspace{2cm} A=r/r_0, \hspace{0.4cm} r_0=(8\pi a
\rho)^{-\frac{1}{2}}, \nonumber \hspace{1.8cm} [45]
\end{equation}
where
\begin{eqnarray}
I_1(A)&=&\int_0^{\infty} \frac{\alpha^2}{1-\alpha^2}\xi d\xi\sin(A\xi),\nonumber\\
I_2(A)&=&\int_0^{\infty} \frac{\alpha}{1-\alpha^2}\xi
d\xi\sin(A\xi),
\end{eqnarray}
and \hspace{1cm}$\alpha=1+\xi^2-\sqrt{(1+\xi^2)^2-1}$. \\We plot
\cite{ass} $D(r)$ for several values of $\sqrt{\rho a^3}$ in
Fig.~\ref{fig:epsart1}.


\begin{references}

\bibitem{NNBogo:47}
N. N. Bogoliubov, J. Phys. J.S.S.R. II, 23 (1947).  \\Bogoliubov
in a footnote thanked Landau for introducing the idea of the
scattering length.

\bibitem{Khuang:87}
K. Huang, Statistical Mechanics, 2nd edition (1987).


\bibitem{LHY}
T. D. Lee, K. Huang and C. N. Yang, Phys. Rev. \textbf{106}, 1135
(1957). Referred to as LHY hereafter. This paper used the
pseudo-potential method: See K. Huang and C. N. Yang, Phys. Rev.
\textbf{105}, 767 (1957); Huang, Yang and Luttinger, Phys. Rev.
\textbf{105}, 776 (1957). The ground state energy $E_0$ was
actually first obtained by a method called binary collision
method:  T. D. Lee and C. N. Yang, Phys. Rev. \textbf{117}, 12
(1960).  Cf. also additional remarks about the binary collision
method by C. N. Yang, p.38-39, p.225-235, Selected Papers
1945-1980 With Commentary.

\bibitem{LPit:98}
L. Pitaevskii and S. Stringari, Phys. Rev. Lett. \textbf{81}, 4541
(1998); G. E. Astrakharchik, R. Combescot, X. Leyronas, and S.
Stringari, Phys. Rev. Lett. \textbf{95}, 030404 (2005); A.
Altmeyer, S. Riedl, C. Kohstall,M. J. Wright, R. Geursen, M.
Bartenstein,C. Chin,J. H. Denschlag and R. Grimm, Phys. Rev. Lett.
\textbf{98}, 040401 (2007); 2008 JILA-Univ. of Colorado preprint;
and 2008 MIT-Harvard preprint.






\bibitem{FLondon}
F. London, Superfluids (1954), pp 142-144 and pp 199-201.

\bibitem{CNYang:62}
C. N. Yang, Rev. Mod. Phys. \textbf{34}, 694 (1962).

\bibitem{ass}
We noticed that equations [46] in LHY are in error. They should
read, for $r\gg r_0$ , $F(r)\rightarrow -G(r)\rightarrow
\frac{1}{4\sqrt{2}}[\pi^2\rho r_0 r^2]^{-1}$. Also [43] in LHY is
in error. It should read as $[43']$ in the present note.

\end{references}
\end{document}